\documentclass[a4paper,10pt]{article}

\usepackage{amsmath}   
\usepackage{amsfonts}
\usepackage{amssymb}

\begin{document}

\setlength{\oddsidemargin} {1cm}
\setlength{\textwidth}{18cm}
\setlength{\textheight}{23cm}

\title{Fully Characterizing Axially Symmetric Szekeres Models With Three Data Sets}

\maketitle

\author{{\bf Marie-No\"elle C\'el\'erier $^{1,a}$, Priti Mishra $^{2,b}$ and Tejinder P. Singh $^{2,c}$} \\
{\small $^1$ Laboratoire Univers et TH\'eories (LUTH), Observatoire de Paris,
CNRS, Universit\'e Paris-Diderot, 5 place Jules Janssen, 92190 Meudon, France} \\
{\small $^2$ Department of Astronomy and Astrophysics, Tata Institute of Fundamental Research, Homi Bhabha Road, Colaba, Mumbai, 400005, Maharashtra, India} \\
{\small e-mail: $^a$ marie-noelle.celerier@obspm.fr} \\
{\small $^b$ priti@tifr.res.in}  \\
{\small $^c$  tpsingh@tifr.res.in \\}}

\begin{abstract}
Inhomogeneous exact solutions of General Relativity with zero cosmological constant have been used in the literature to challenge the $\Lambda$CDM model. From one patch Lema\^itre-Tolman-Bondi (LTB) models to axially symmetric quasi-spherical Szekeres (QSS) Swiss-cheese models, some of them are able to reproduce to a good accuracy the cosmological data. It has been shown in the literature that a zero $\Lambda$ LTB model with a central observer can be fully determined by two data sets. We demonstrate that an axially symmetric zero $\Lambda$ QSS model with an observer located at the origin can be fully reconstructed from three data sets, number counts, luminosity distance and redshift drift. This is a first step towards a future demonstration involving five data sets and the most general Szekeres model.
\end{abstract}

\section{The Szekeres solution}

We are interested here in finding an algorithm for building an axially symmetric quasi-spherical Szekeres (QSS) model from a set of observables. The problem to be solved is to determine the number of observables needed and to give a recipe allowing us to fully define the model from them. This problem has already been solved for Lema\^itre-Tolman-Bondi models with a central observer by Mustapha et al. \cite{MHE97}.

The QSS line element in comoving and synchronous coordinates is
\begin{equation}
{\rm d} s^2 = {\rm d} t^2 - \frac{(\Phi' - \Phi E'/ E)^2}{
1 - k} \, {\rm d} r^2 - \frac{\Phi^2}{E^2} ({\rm d} x^2 + {\rm d} y^2),
\label{metric}
\end{equation}
where $\Phi$ is a function of the $t$ and $r$ coordinates and
\begin{equation}
E = \frac{S}{2} \left[ \left(\frac{x-P}{S}\right)^2 + \left(\frac{y-Q}{S}\right)^2  + 1 \right].
\label{Edef}
\end{equation}
The Einstein equations give
\begin{equation}
\dot{\Phi}^2 = \frac{2M}{\Phi} - k + \frac{1}{3} \Lambda \Phi^2,
\label{evoleq}
\end{equation}
\begin{equation}
\kappa \rho = \frac{2M' - 6M E'/ E}{\Phi^2 ( \Phi' - \Phi E'/ E)},
\label{rho}
\end{equation}
where a dot denotes differentiation with respect to $t$ and a prime, differentiation with respect to $r$. Equation (\ref{evoleq}) can be integrated as
\begin{equation}
\int\limits_0^{\Phi} \frac{{\rm d} \widetilde{\Phi}}{\sqrt{2M /
\widetilde{\Phi} - k +  (1 / 3) \Lambda \widetilde{\Phi}^2}} = t - t_B.
\label{tB}
\end{equation}
Here, $S(r), P(r), Q(r), k(r), M(r), t_B(r)$ are all functions of the $r$ coordinate.

\section{Set of differential equations}

In the axially symmetric case, we can set $P=Q=0$ \cite{ND07}. The model is thus defined by three independent functions of $r$, plus a coordinate choice. We use this freedom to define the $r$ coordinate such that, on the past light cone of the observer located at $(t= t_0, r=0)$,
\begin{equation}
\frac{\widehat{\Phi'} - \widehat{\Phi} E'/E}{\sqrt{1 - k}} = 1,
\label{coordch}
\end{equation}
where a hat denotes quantities computed on the past light cone. Substituting our coordinate choice into the total derivative of $\widehat{\Phi}$ and using Eq.~(\ref{evoleq}) with $\Lambda = 0$, we obtain
\begin{equation}
\frac{E'}{E} = \frac{1}{\widehat{\Phi}} \left(\frac{{\rm d} \widehat{\Phi}}{{\rm d} r} + \sqrt{\frac{2 M}{\widehat{\Phi}} -k} - \sqrt{1 - k}\right).
\label{E'E}
\end{equation}
Now, we substitute Eq.~(\ref{E'E}) into Eq.~(\ref{rho}), using Eq.~(\ref{coordch}), which gives
\begin{equation}
\frac{{\rm d}M}{{\rm d} r} - \frac{3M}{\widehat{\Phi}} \left[\frac{{\rm d} \widehat{\Phi}}{{\rm d} r} + \sqrt{\frac{2M}{\widehat{\Phi}} -k} - \sqrt{1 - k}\right]
- 4 \pi \widehat{\rho} \widehat{\Phi}^2 \sqrt{1 - k} = 0.
\label{dMdr}
\end{equation}
Then, we consider the luminosity distance on the axially directed past null geodesic, $\widehat{D_L}(z) = \widehat{D_A}(z)\left(1 + \widehat{z}\right)^2$ which we substitute into the equation for $\widehat{D_A}$ as given by Bolejko and C\'el\'erier \cite{BC10} and obtain a differential equation of the form
\begin{equation}
F\left({\rm d^2} \widehat{D_L} / {\rm d} z^2, {\rm d} \widehat{D_L} / {\rm d} z, \widehat{D_L}, {\rm d^2} \widehat{\Phi} / {\rm d} r^2, {\rm d} \widehat{\Phi} / {\rm d} r, \widehat{\Phi}, {\rm d} \widehat{\rho} / {\rm d} z, \widehat{\rho}, \widehat{z}, M, k, k', k''\right) = 0,
\label{DL}
\end{equation}
The third differential equation needed proceeds from the equation for the redshift and is of the form
\begin{equation}
G\left({\rm d} \widehat{\Phi} / {\rm d} r, \widehat{\Phi}, {\rm d} \widehat{z} / {\rm d} r, \widehat{z}, \widehat{\rho}, M, k, k'\right) = 0.
\label{redshift}
\end{equation}

We have obtained a set of three coupled differential equations for the unknown functions $M, k$, and $\widehat{\Phi}$, in which $ \widehat{z}$ and $\widehat{D_L}, \widehat{\rho}$ and their derivatives with respect to $z$ are observables. Now, to be able to determine the model, we need $M, k$, and $\widehat{\Phi}$ as functions of $r$. Since the data are given in terms of the redshift $z$, we have to transform them into functions of $r$. We use thus the redshift drift, $\widehat{\dot{z_0}}(z) = (\delta \widehat{z}/ \delta t_0)(z)$ as given by Mishra et al. \cite{MCS12} and obtain a differential equation for $\widehat{z}(r)$:
\begin{equation}
\frac{{\rm d^2} \widehat{z}}{{\rm d} r^2} - \frac{1}{1 + \widehat{z}} \left(\frac{{\rm d} \widehat{z}}{{\rm d} r}\right)^2 + \left[ \left(1 + \widehat{z}\right) \frac{{\rm d} \widehat{\dot{z_0}}}{{\rm d} z} - \widehat{\dot{z_0}}\right]\frac{{\rm d} \widehat{z}}{{\rm d} r} = 0.
\label{ddrift}
\end{equation}
We solve numerically the above set of differential equations with the origin conditions $M(r=0) = k(r=0) = \widehat{\Phi}(r=0) = \widehat{z}(r=0) = 0$,  and obtain $M(r), k(r)$ then $(E'/E)(r)$ and thus $S(r)$, from the relation $E'/E = S'/S$, valid on an axially directed geodesic. With $P(r) = Q(r) = 0$ and our coordinate choice, this is sufficient to fully determine the model.

\section{Algorithm}

To obtain the functions, $M$, $k$, $E$ and $t_B$ from the galaxy number count, $n(z)$, supernova luminosity distance and redshift drift observations, we propose the following algorithm:

\begin{enumerate}

\item Make a rigorous processing of the data to obtain smooth analytic functions of $\widehat{D_L}(z), \widehat{\dot{z}_0}(z)$ and $\widehat{\rho}(z) = m n(z)$, assumed to obtain on the axial null geodesic directed towards the observer located at the origin.

\item Determine $\widehat{z}(r)$ by solving numerically Eq.~(\ref{ddrift}).

\item Determine $\widehat{\Phi}(r)$, $M(r)$ and $k(r)$ from the set of three coupled differential equations (\ref{dMdr})-(\ref{redshift}).

\item Knowing $M(r), k(r)$ and $\widehat{\Phi}(r)$, determine $E(r)$ (equivalently $S(r)$) by integrating Eq.~(\ref{E'E}).

\item Knowing now the sign of $E$, $t_B(r)$ follows from the resolution of the parametric equations issued from Eq.~(\ref{tB}) \cite{BKHC10}.

\end{enumerate}

\section{Conclusion}

We have given here a set of equations and an algorithm able to fully determine an axially symmetric QSS model of universe from three data sets assumed to be measured on an axially directed null geodesic issued from an observer located at the origin. This algorithm can only be applied to an {\it ideal} universe, where the redshift is monotonically increasing and where no shell-crossing is present to close the past null cone. However, since axial symmetry is not a very credible feature of our observed Universe, we do not claim this work to be more than a mere exercise designed to go beyond spherical symmetry and which will have to be generalized in the future to more physical Szekeres models.

\end{document}